put
\documentstyle[prl,aps,twocolumn]{revtex}

\begin{document}

\title{ Einstein's equations in Ashtekar's variables constitute\\
 a symmetric hyperbolic system.}

\author{Mirta S. Iriondo\thanks{Supported by STINT, The Swedish Foundation for International Cooperation in Research and Higher Education. }
 , Enzo O. Leguizam\'on
\thanks{Fellow of Se.CyT-UNC.}
\and and Oscar A. Reula
\thanks{Member of CONICET.}\\
{\small FaMAF, Medina Allende y Haya de la Torre,}\\
 {\small Ciudad Universitaria, 5000 C\'ordoba, Argentina}\\
{\small Fax nr: +54-51-334054}\\}

\maketitle

\vspace{-.4in}
\begin{abstract}
We show that the $3+1$ vacuum Einstein field equations in Ashtekar's variables constitutes a first order symmetric hyperbolic system for arbitrary but fixed lapse and shift fields,  by suitable adding to the system terms proportional to the constraint equations.
\end{abstract}
\pacs{...}
\section{Introduction}
\label{sec:int}

In the mid-eighties a new representation for Hamiltonian General Relativity was introduced  by Ashtekar \cite{abhay1}. The Ashtekar representation maybe understood as resulting from a complex canonical transformation that goes from the ADM gravitational phase space variables to a new set of variables. One of the major benefits of this representation is that the constraints take a simpler form in terms of these variables than in the ADM formulation. Another important aspect of this set of variables, surprisingly not yet exploited and not shared by the standard ADM representation, is that the evolution equation are a  first order system of differential equations. Thus it is natural to consider the well-posedness of the classical initial value problem in this context.

Well posedness of Einstein equations in itself is a long ago closed issue. Since the initial data for the equations in Ashtekar's variables is in one to one, and continuous correspondence with the ADM variables, the result follows at once. Rather, the issue we pose here is whether Einstein's equations in Ashtekar's variables conform a good system to evolve arbitrary initial data using approximations schemes, be numerical or analytical. Bad choices of evolution gauges usually lead into spurious coordinate singularities which effectively destroy the smoothness of the map between variables. We are also concerned with consistency, convergence and stability properties of approximation methods which  use  these variables. These issues have not been fully solved in the realm of the numerical analysis of non linear partial differential evolution equations, but all powerful results so far obtained make heavy use of the symmetric hyperbolic character of the equations. This is the main reason for checking this property
 here.  Updated references to this problem are in \cite{helmut1}, \cite{choquet}, \cite{Bona}, \cite{Anninos}, and \cite{reula}.

As discussed in the conclusions (section \ref{sec:conc}),  one important property of the system we consider is that the lapse-shift pair can be given before hand on spacetime or can be obtained along evolution, either imposing evolution equations and so effectively enlarging the symmetric hyperbolic system, or imposing elliptic equations on each hypersurface $t=const$.

The plan of the paper is as follows. Section \ref{sec:ash} is devoted to mathematical preliminaries where we recall some aspects of  Ashtekar's Hamiltonian formulation. In section \ref{sec:sym} we present the main result.

\section{The field equations in Ashtekar's variables}
\label{sec:ash}
In this section we shall present the necessary definitions and a brief overview of the Hamiltonian formalism in spinorial variables in order to get Einstein field equations as a system of evolution and constraint equations (we  follow  \cite{abhay}, \cite{sen}, and \cite{smolin}).

Let $M$ be a four dimensional manifold with a smooth metric $g_{ab}$ of signature $(-+++)$. Furthermore, assume that $M$ admits a time function $t$, a smooth scalar field on $M$, whose gradient is  everywhere timelike and  denote the time direction  by $t^a$,  which is a smooth vector field on $M$ with affine parameter $t$, i.e. $t^a\nabla_a t=1$. Then the one parameter family of surfaces $\Sigma_t$, defined by $t=const.$, are three dimensional spacelike surfaces. Finally  we shall denote by $n^a$ the future directed unit normal  vector field  to $\Sigma_t$, and  the intrinsic and  positive definite metric on $\Sigma_t$ by $q_{ab}$. By projecting $t^a$ into and orthogonally to $\Sigma_t$, we obtain the lapse and the shift fields $N$ and $N^a$, thus $t^a=Nn^a+N^a$.
 
We shall also assume that our spacetime  admits an spinorial structure (see  \cite{penrose}, and references therein ); i.e we shall assume that  to each point $p \in M$ we can assign a complex two dimensional vector space $W$ equipped with a nondegenerated symplectic form (skew 2-form). The elements of $W$ will be denoted by $\xi^A$ and the corresponding symplectic form by $\epsilon_{AB}$ (its inverse will be denoted by $\epsilon^{AB}$, and is such that $\epsilon_{AC}\epsilon^{AB}=\delta_C{}^B$). The complex conjugate vector space associated with $W$ will be denoted by $\overline{W}$, its elements by  $\overline{\xi}^{A' }$ and the symplectic form by $\epsilon_{ A' B'}$\footnote{For more definitions see \cite{sen}.}. Since the group that preserves the structure on $W$ is $SL(2,C)$,  $\xi^A$ will be called a $SL(2,C)$ spinor at $p\in M$. 

To relate tensors to spinors on $M$, one must fix a metric preserving isomorphism between the four dimensional real vector space $V=\{ \xi^{A A'}\in W \otimes \overline{W}: \overline{\xi^{A A'}}=-\xi^{A A'}, g_{AA'BB'}=\epsilon_{AB}\epsilon_{A'B'}\}$  and the tangent space $T_p(M)$ with metric $g_{ab}$. This isomorphism is called an  $SL(2,C)$ soldering form and is denoted by $ \sigma^a{}_{A A'}$, Then 
$$
g^{ab}=\sigma^a{}_{A A'}\sigma^b{}_{B B'}\epsilon^{AB}\epsilon^{A' B'}.
$$
The covariant derivative operator on tensors has a unique extension to spinors if we impose the condition
$$
\nabla_a \sigma^b{}_{AA'}=0.
$$
In order to relate  spinors on these surfaces to their  intrinsic geometry, we shall need an additional structure on $W$ which reduces the structure group $SL(2,C)$ to $SU(2)$. This additional structure on $W$ is a positive definite hermitian inner product which is denoted by $G: W\times W \to C$, or 
$$
G_{AA'}: W\times \overline{W} \to C, \qquad (\eta^A,\overline{\xi}^{A'})\mapsto \overline{\xi}^{A'} G_{A'A}\eta^A.
$$
In this notation hermiticity appears as $G_{A'A}=\overline{G}_{AA'}$. We will assume that (either $\epsilon_{AB}$ or) $G_{AA'}$ is so normalized that
$$
\epsilon^{A'B'}G_{A'A}G_{B'B}=\epsilon_{AB}, \quad \mbox{or} \qquad G_{A'A}G^{A'B}=\delta_A{}^B,
$$
In the formalism $3+1$ this additional structure is obtained by choosing
$$
G_{AA'}:=\sqrt{2} \; i \;n_{AA'}:=\sqrt{2} \; i \;n^a\;\sigma_a{}_{AA'},
$$
then as $\overline{\sigma}_a{}^{AA'}=-\sigma_a{}^{AA'}$ and $n^a$ is timelike and future directed, we have the positive definite  hermitian inner product that we were looking for. Via this inner product, we identify unprimed $SL(2,C)$ spinors on $M$ with $SU(2)$ spinors on $\Sigma_t$. The relationship between the intrinsic geometry with  $SU(2)$ spinors is given by:
$$
\sigma_a{}^A{}_B:=q_a{}^b\sigma_b{}^{AA'}G_{A'B}
$$
that is the $SU(2)$ soldering form. Then $$
q_{ab}=-\sigma_a{}_A{}^B\sigma_b{}_B{}^A:=-\mbox{tr}(\sigma_a\sigma_b).
$$
Here a matrix notation is employed (and shall be used in the following) for unprimed spinor indices in which adjacent summed indices go from upper left to lower right, e.g., $(\sigma^a\sigma^b)_A{}^B=\sigma^a{}_A{}^C\sigma^b_C{}^B$.

There is a natural (canonical) spinorial connection associated with the metric such that $D_a \sigma^b=0$. The pullback of the 4-dimensional canonical connection to the hypersurface is another connection on the 3-surface called {\it Sen Connection}, defined by
$$
{\cal D}_a\alpha_A=q_a{}^b\nabla_b\alpha_A,
$$
the curvature associated to this connection is denoted by $F_{ab}{}_A{}^B$ and is the pullback of the 4-dimensional curvature $F_{ab}{}_A{}^B=q_a{}^cq_b{}^d\;{}^4R_{cd}{}_A{}^B$. If we fix (for simplicity) a connection $\partial$ on the unprimed spinor indices, chosen to be real and flat,  and replace $\nabla$ by a $SL(2,C)$ Lie algebra-valued connection 1-form ${}^4A_{aA}{}^B$:
$$
\nabla_a \lambda_A:= \partial_a \lambda_A+ {}^4A_{aA}{}^B\lambda_B,
$$
 the curvature tensor becomes
$$
{}^4R_{ab}:=2\partial_{[a}{}^4A_{b]}+[{}^4A_a,{}^4A_b].
$$
Recall that ${}^4R_{ab}{}_A{}^B$ is defined by
$$
(\nabla_a\nabla_b-\nabla_b\nabla_a)\lambda_A={}^4R_{abA}{}^B\lambda_B,
$$
thus we can write $F_{ab}{}_C{}^D$ in terms of a $SU(2)$ Lie algebra valued one form $A_a{}_C{}^D$ 
$$
F_{ab}:=2\partial_{[a}A_{b]}+[A_a,A_b].
$$
Using  these  variables the action can be defined as:
$$
S(\sigma,{}^4A)=\int \mbox{d}^4 x \;({}^4\sigma)\sigma^a{}_A{}^{A'} \sigma^b{}_{BA'} {}^4R_{ab}{}^{AB},
$$
where $({}^4\sigma)$ denotes the square root of minus the determinant of the four metric,  this action is presented in \cite{smolin} and in \cite{abhay}.
  Using the identity:
\begin{eqnarray*}
\sigma^a{}_A{}^{A'}\sigma^b{}_{BA'}{}^4R_{ab}{}^{AB}&=&\;\mbox{tr}\bigg(\sqrt{2}\; i\; n^a\sigma^b {}^4R_{ab}-\sigma^a\sigma^b{}^4R_{ab}\bigg)\\
&=&\;\mbox{tr}\bigg(-\sigma^a\sigma^b{}^4R_{ab}+\sqrt{2}\; i \;N^{-1}\sigma^b\\
&&\times({\cal L}_t{}^4A_b-{\cal D}_b({}^4A\cdot t))\\
&&-\sqrt{2}\; i \;N^{-1}N^a\sigma^b{}^4R_{ab}\bigg),
\end{eqnarray*}
where ${}^4A\cdot t={}^4A_at^a$,  we have
\begin{eqnarray*}
S&=&\int \mbox{dt}\int\mbox{d}^3x({}^4\sigma)\;\mbox{tr}\bigg(\sqrt{2}\; i N^{-1}\sigma^b({\cal L}_t{}^4A_b\\
&&-{\cal D}_b({}^4A\cdot t))-\sigma^a\sigma^b{}^4R_{ab}-\sqrt{2}\; i N^{-1}N^a\sigma^b{}^4R_{ab}\bigg).
\end{eqnarray*}
Finally, since  $\sigma^a{}_{AB}$ projects onto the 3-surface, and ${}^4\sigma=N\sigma$,  the action\footnote{The surface terms  are not included here.} becomes 
\begin{eqnarray*}
S&=&\int \mbox{dt}\int\mbox{d}^3x\;\mbox{tr}\bigg(\sqrt{2}\; i \tilde\sigma^b({\cal L}_t A_b-{\cal D}_b({}^4A\cdot t))\\
&&-\;\hbox {${}_{{}_{{}_{\widetilde{}}}}$\kern-.5em\rm N}\;\;\tilde\sigma^a\tilde\sigma^b F_{ab}-\sqrt{2}\; i N^a\tilde\sigma^b F_{ab}\bigg).
\end{eqnarray*}
with $\tilde\sigma^a{}_{AB}=(\sigma)\;\sigma^a{}_{AB}$ and $\;\hbox {${}_{{}_{{}_{\widetilde{}}}}$\kern-.5em\rm N}=(\sigma)^{-1}\; N$.

 The action depends on  five  variables $\;\hbox {${}_{{}_{{}_{\widetilde{}}}}$\kern-.5em\rm N}$, $N^a$, ${}^4A\cdot t$, $A_{aA}{}^B$ and $\tilde\sigma^a{}_{AB}$. The first three variables play the role of the Lagrange multipliers, only the last two are dynamical variables.
 Varying the action with respect to the Lagrange multipliers we obtain the constraint equations:
\begin{eqnarray}
C(\tilde\sigma,A)&:=&\mbox{tr}(\tilde\sigma^a\tilde\sigma^bF_{ab})=0,\nonumber\\
C_a(\tilde\sigma,A)&:=&\mbox{tr}(\tilde\sigma^bF_{ab})=0,\label{eq:constraint}
\\
\tilde C_A{}^B(\tilde\sigma,A)&:=&{\cal D}_a \tilde\sigma^a{}_A{}^B=0.\nonumber
\end{eqnarray}
varying with respect to the dynamical variables,  yields the evolution equations:
\begin{eqnarray}
\label{eq:evolution}
{\cal L}_t\tilde\sigma^b&=&[{}^4A\cdot t,\tilde\sigma^b]+2{\cal D}_a(N^{[a}\tilde\sigma^{b]})\nonumber\\
&&-\frac{1}{\sqrt{2}}i{\cal D}_a(\;\hbox {${}_{{}_{{}_{\widetilde{}}}}$\kern-.5em\rm N}[\tilde\sigma^{a},\tilde\sigma^{b}]),\nonumber\\
{\cal L}_t A_b&=&{\cal D}_b({}^4A\cdot t)+N^aF_{ab}+\frac{1}{\sqrt{2}}i\;\;\hbox {${}_{{}_{{}_{\widetilde{}}}}$\kern-.5em\rm N}[\tilde\sigma^a,F_{ba}].
\end{eqnarray}

\section{Symmetric Hyperbolicity}
\label{sec:sym}

In order to obtain a symmetric hyperbolic system for any given lapse-shift pair,   we suitably modify the field equations outside the constraint submanifold. Indeed, the modification only involves the addition of terms proportional to the constraints. Since we know that the evolution vector field is tangent to this manifold, the physical evolution, that is  the dynamics inside the constraint manifold,  remains unchanged.  The modified  evolution equations are 
 
\begin{eqnarray}
\label{eq:evolution1}
{\cal L}_t\tilde\sigma^b&=&[{}^4A\cdot t,\tilde\sigma^b]+2 {\cal D}_a(N^{[a}\tilde\sigma^{b]})\nonumber\\
&&-\frac{1}{\sqrt{2}}i{\cal D}_a(\;\hbox {${}_{{}_{{}_{\widetilde{}}}}$\kern-.5em\rm N}[\tilde\sigma^{a},\tilde\sigma^{b}])+\frac{1}{\sqrt{2}}i\;\;\hbox {${}_{{}_{{}_{\widetilde{}}}}$\kern-.5em\rm N} [\tilde C,\tilde\sigma^b]+N^b \tilde C,\nonumber\\
{\cal L}_t A_b&=&{\cal D}_b({}^4A\cdot t)+N^aF_{ab}+\frac{1}{\sqrt{2}}i\;\;\hbox {${}_{{}_{{}_{\widetilde{}}}}$\kern-.5em\rm N}[\tilde\sigma^a,F_{ba}]\nonumber\\
&&+\frac{i}{\sigma^2\sqrt{2}}\;\;\hbox {${}_{{}_{{}_{\widetilde{}}}}$\kern-.5em\rm N} \tilde\sigma_b C+\frac{i}{\sigma^4}\;\;\hbox {${}_{{}_{{}_{\widetilde{}}}}$\kern-.5em\rm N} \tilde\epsilon_b{}^{dc}\tilde\sigma_c C_d.
\end{eqnarray}
Since the Lagrange multiplier ${}^4A_a t^a$  is an arbitrary function we can choose it as ${}^4A_a t^a=A_a N^a$, i.e.  we set its normal projection to zero.  In this way, in the equation for the evolution of $A_a$,  the principal part corresponding to the first two terms is replaced by the directional derivative of $A_a$. 

To see that the  system (\ref{eq:evolution1}) is symmetric hyperbolic, we shall only need to consider its principal symbol. Recall that the principal  symbol of a quasilinear evolution equation system
$$
\dot u^l=B^l{}_j{}^a(u)\nabla_a u^j + M^l(u)
$$
is given by $P(u,ik_a)=i\;B^l{}_j{}^a(u) k_a$. In our case $u$ denotes $u=(\sigma^a,A_b)$.

{\bf Lemma III.1:} {\em The equation system (\ref{eq:evolution1}) for any fixed, but arbitrary lapse and shift fields is a symmetric hyperbolic system}.

\noindent{{\bf Proof:}}

Since the vector $u$ in our case is complex, symmetry of real systems corresponds to antihermiticity of the principal symbol with respect to some local hermitian inner product, i.e.  an hermitian, positive definite bilinear form. In our case the natural one:
\begin{eqnarray*}
\langle u_2, u_1\rangle &\equiv &\langle(\tilde\sigma_2,A^2),(\tilde\sigma_1,A^1)\rangle\\
&\equiv & \mbox{tr}(\tilde\sigma^{\dag}_2{}^a \tilde\sigma_1^b)\;q_{ab}+  \mbox{tr}(A^{2\dag}{}_a A^{1}{}_b)\;q^{ab},
\end{eqnarray*}
is the one that works. Thus we need to prove that
$$
P_{12}+P^{\dag}{}_{12}\equiv \langle u_2,Pu_1\rangle + \langle u_2,P^{\dag}  u_1\rangle=0.
$$
To obtain an expression for the symbol $Pu_1$ we neglect all the terms which do not have derivatives of the dynamical variables $\tilde \sigma^b$ and $A_b$. We remark  the {\em very important} fact that for both equations, (\ref{eq:evolution}) and (\ref{eq:evolution1}),  the principal part decouples into two blocks one acting on $\sigma^b$ and the other acting on $A_a$, so we can write 
\begin{equation}
\label{eq:ps}
P(\;\hbox {${}_{{}_o}$\kern-.6em\rm u},ik_a)=P_\sigma(\;\hbox {${}_{{}_o}$\kern-.6em\rm u},ik_a) \oplus P_A(\;\hbox {${}_{{}_o}$\kern-.6em\rm u},ik_a),
\end{equation}
in our case $\hbox {${}_{{}_o}$\kern-.6em\rm u}\;\;=(\tilde\sigma,A)$. The action of $P$ on any solution $u$ can be written as
\begin{eqnarray}
\label{eq:op}
P_\sigma(\tilde\sigma,ik_a)\tilde\sigma_1{}^b&=&\frac{k_a}{\sqrt{2}}\;\;\hbox {${}_{{}_{{}_{\widetilde{}}}}$\kern-.5em\rm N}[\tilde\sigma^a,\tilde\sigma_1{}^b]+i\;k_a N^a \tilde\sigma_1{}^b,\nonumber\\
P_A(\tilde\sigma,ik_c)\;A^1{}_a&=&i\;N^b\;k_b A^1{}_{a}-\frac{1}{\sqrt{2}}\;\hbox {${}_{{}_{{}_{\widetilde{}}}}$\kern-.5em\rm N}\;\;[\tilde\sigma^b,k_aA^1{}_{b}]\nonumber\\
&&+\frac{1}{\sqrt{2}}\;\hbox {${}_{{}_{{}_{\widetilde{}}}}$\kern-.5em\rm N}\;\;[\tilde\sigma^b,k_bA^1{}_{a}]\nonumber\\
&&-\frac{1}{\sqrt{2}}\;\hbox {${}_{{}_{{}_{\widetilde{}}}}$\kern-.5em\rm N}\;\;\tilde\sigma_a\; \mbox{tr} ( [\tilde\sigma^d,\tilde\sigma^b]k_dA^1{}_{b})\\ &&-\;\hbox {${}_{{}_{{}_{\widetilde{}}}}$\kern-.5em\rm N}\;\;\tilde\epsilon_a{}^{bc}\tilde\sigma_c\;\mbox{tr} (\tilde\sigma^dk_bA^1{}_{d})\nonumber\\
&&+\;\hbox {${}_{{}_{{}_{\widetilde{}}}}$\kern-.5em\rm N}\;\;\tilde\epsilon_a{}^{bc}\tilde\sigma_c\;\mbox{tr} (\tilde\sigma^dk_dA^1{}_{b}) .\nonumber
\end{eqnarray}
Adding the expressions for $P_\sigma{}_{12}$ and $P_\sigma^{\dag}{}_{12}$ we obtain

\begin{eqnarray}
\label{eq:ops}
P_\sigma{}_{12}+P_\sigma^{\dag}{}_{12}&\equiv &\langle \tilde\sigma_2,P_\sigma \tilde\sigma_1\rangle +\langle \tilde\sigma_2,P^{\dag}{}_\sigma \tilde\sigma_1\rangle \nonumber\\
&=&-\frac{k_a}{\sqrt{2}}\;\;\hbox {${}_{{}_{{}_{\widetilde{}}}}$\kern-.5em\rm N}\mbox{tr}(\tilde\sigma^a[\tilde\sigma_1{}^b,\tilde\sigma_2{}_b])\\
&&+\frac{k_a}{\sqrt{2}}\;\;\hbox {${}_{{}_{{}_{\widetilde{}}}}$\kern-.5em\rm N}\mbox{tr}(\tilde\sigma^a[\tilde\sigma_1{}_b,\tilde\sigma_2{}^b])\nonumber\\
&&-i k_a N^a\mbox{tr}(\tilde\sigma_2{}_b\tilde\sigma_1{}^b)+i k_a N^a\mbox{tr}(\tilde\sigma_2{}^b\tilde\sigma_1{}_b)\nonumber\\
&=&0.\nonumber
\end{eqnarray}
here we have used that $\tilde\sigma^{a\dag}=-\tilde\sigma^a$.   It only remains to prove a similar result for $P_A{}_{12}$, using in the expression for $P_A A^1{}_a$ that
$$
A^1{}_a=-\frac{1}{\sigma^2}\mbox{tr} ( A^1{}_a\tilde\sigma_e)\tilde\sigma^e
$$
and
$$
[\tilde\sigma^a,\tilde\sigma^b]_A{}^B=\frac{\sqrt{2}}{\sigma^2}\tilde\epsilon^{ab}{}_{c}\tilde\sigma^c{}_A{}^B,
$$
we obtain

\begin{eqnarray*}
\frac{\sigma^4}{\;\hbox {${}_{{}_{{}_{\widetilde{}}}}$\kern-.5em\rm N}}P_A A^1{}_a&=&\tilde\epsilon^{bem}\tilde\sigma_mk_a\;\mbox{tr} ( A^1{}_b\tilde\sigma_e)-\frac{\sigma^2\;i}{\;\hbox {${}_{{}_{{}_{\widetilde{}}}}$\kern-.5em\rm N}}N^b\;k_b \mbox{tr}(A^1{}_{a}\tilde\sigma_e)\tilde\sigma^e
\\
&&-\tilde\epsilon^{bem}\tilde\sigma_mk_b\;\mbox{tr} ( A^1{}_a\tilde\sigma_e)-\tilde\epsilon^{dbe}\tilde\sigma_ak_d\;\mbox{tr} ( A^1{}_b\tilde\sigma_e)\\
&&-\tilde\epsilon_a{}^{bc}\tilde\sigma_ck_b\;\mbox{tr} ( A^1{}_d\tilde\sigma^d)+\tilde\epsilon_a{}^{bc}\tilde\sigma_ck_d\;\mbox{tr} ( A^1{}_b\tilde\sigma^d).
\end{eqnarray*}
Thus we have 
\begin{eqnarray}
\label{eq:opa}
\frac{\sigma^4}{\;\hbox {${}_{{}_{{}_{\widetilde{}}}}$\kern-.5em\rm N}}(P_A{}_{12}+P^{\dag}{}_A{}_{12})&=&-\frac{\sigma^2\;i}{\;\hbox {${}_{{}_{{}_{\widetilde{}}}}$\kern-.5em\rm N}}\;N^b\;k_b \mbox{tr}(A^{2\dag}{}^a\tilde\sigma^e)\mbox{tr}(A^1{}_{a}\tilde\sigma_e)\nonumber\\
&&+\frac{\sigma^2\;i}{\;\hbox {${}_{{}_{{}_{\widetilde{}}}}$\kern-.5em\rm N}}\;N^b\;k_b \mbox{tr}(A^{2\dag}{}_a\tilde\sigma^e)\mbox{tr}(A^1{}^{a}\tilde\sigma_e)\nonumber\\
&&+\tilde\epsilon^{bem}\;\mbox{tr}(A^{2\dag}{}^a\tilde\sigma_m)k_a\;\mbox{tr} ( A^1{}_b\tilde\sigma_e)\nonumber\\
&&-\tilde\epsilon^{bem}\;\mbox{tr}(A^{2\dag}{}^a\tilde\sigma_m)k_b\;\mbox{tr} ( A^1{}_a\tilde\sigma_e)\nonumber\\
&&-\tilde\epsilon^{dbe}\;\mbox{tr}(A^{2\dag}{}^a\tilde\sigma_a)k_d\;\mbox{tr} ( A^1{}_b\tilde\sigma_e)\nonumber\\
&&-\tilde\epsilon_a{}^{bc}\;\mbox{tr}(A^{2\dag}{}^a\tilde\sigma_c)k_b\;\mbox{tr} ( A^1{}_d\tilde\sigma^d)\nonumber\\
&&+\tilde\epsilon_a{}^{bc}\;\mbox{tr}(A^{2\dag}{}^a\tilde\sigma_c)k_d\;\mbox{tr} ( A^1{}_b\tilde\sigma^d)\\
&&+\tilde\epsilon^{bem}\;\mbox{tr}(A^{2\dag}{}_b\tilde\sigma_e)k_a\;\mbox{tr} ( A^1{}^a\tilde\sigma_m)\nonumber\\
&&-\tilde\epsilon^{bem}\;\mbox{tr}(A^{2\dag}{}_a\tilde\sigma_e)k_b\;\mbox{tr} ( A^1{}^a\tilde\sigma_m)\nonumber\\
&&-\tilde\epsilon^{dbe}\;\mbox{tr}(A^{2\dag}{}_b\tilde\sigma_e)k_d\;\mbox{tr} ( A^1{}^a\tilde\sigma_a)\nonumber\\
&&-\tilde\epsilon_a{}^{bc}\;\mbox{tr}(A^{2\dag}{}^d\tilde\sigma_d)k_b\;\mbox{tr} ( A^1{}^a\tilde\sigma_c)\nonumber\\
&&+\tilde\epsilon_a{}^{bc}\;\mbox{tr}(A^{2\dag}{}_b\tilde\sigma^d)k_d\;\mbox{tr} ( A^1{}^a\tilde\sigma_c)\nonumber.
\end{eqnarray}
Rearranging the indices and using the antisymmetry of the Levi-Civita tensor density we get that the first term cancels out  the second,  the $4th$  cancells out the $9th$ , and  the $5th$ and $6th$ cancel $10th$ and $11th$.
 
Finally, using 
 
$$
2A^{[a}\tilde\sigma^{b]}=\frac{1}{\sigma^2}\tilde \epsilon^{abe}\tilde \epsilon_{dce}A^d\tilde\sigma^c,
$$
the remaining terms vanish,  yielding  the antihermiticity of the principal symbol. This concludes the proof of the Lemma.$\spadesuit$

\section{Concluding Remarks}
\label{sec:conc}

The result obtained here assumes,  for simplicity,  a given and  fixed,  lapse-shift pair. This assumption is not very practical, for one would like, for instance,  to choose the pair along the evolution in order to avoid the occurrence of spurious coordinate singularities. There are at least two ways to loosen the assumption made. The first is to incorporate the lapse-shift pair into the evolution equations by choosing appropriate equations for them in such a way as to obtain a bigger symmetric hyperbolic system. The second is to impose some elliptic equations on the pair in such a way as to be able to estimate the $L^2$ norm of the derivatives of it -for they appear on the equations for $(\tilde\sigma,A)$- in terms of the $L^2$ norms of $(\tilde\sigma,A)$. This would be the case if one were to use Witten's equation to determine the pair. Besides these two ways of fixing the lapse-shift pairs along evolution there does not seem to be any other obvious candidate.

\end{document}